\documentclass[11pt]{article}
\newcommand{\One}{{\hbox{{\rm 1{\hbox to 1.5pt{\hss\rm1}}}}}}
\newcommand{\half}{{\textstyle\frac{1}{2}}}
\newcommand{\vev}[1]{\langle\,#1\,\rangle}

\newcommand{\CM}{{\cal M}}

\newcommand{\CO}{{\cal O}}
\newcommand{\CP}{{\cal P}}
\newcommand{\CQ}{{\cal Q}}

\newcommand{\CZ}{{\cal Z}}

\newcommand\be{\begin{equation}}
\newcommand\ee{\end{equation}}
\newcommand\bea{\begin{eqnarray}}
\newcommand\eea{\end{eqnarray}}
\newcommand\nn{\nonumber}
\newcommand\ba{\(\begin{array}}
\newcommand\ea{\end{array}\)}
\newcommand{\resection}[1]{\setcounter{equation}{0}\section{#1}}

\begin{document}
\vskip 1.5cm
\begin{center}
{\Large{\bf
Bulk one-point function on disk in one-matrix model
}}
\end{center}
\vskip 1cm
\centerline{Alexander Belavin%
\footnote{e-mails: belavin@itp.ac.ru}
and 
Chaiho Rim%
\footnote{e-mail: rimpine@sogang.ac.kr}}
\vskip 0.9cm
\centerline{${}^1$\sl\small L.D. Landau Institute for 
Theoretical Physics, Chernogolovka 142432, Russia}
\vskip 0.2cm
\centerline{${}^2$\sl\small 
Dept. of Physics and Center for Quantum Spacetime, 
Sogang University,
Seoul  121-742, Korea}
\vskip 1.25cm
\begin{abstract}
\noindent
We consider bulk correlation numbers on disk
in one-matrix model. Using the recently found
so-called resonance transformation from 
the KdV to the Liouville
frame\cite{BZ}, 
we obtain an explicit expression for the bulk
one-point function. The result is consistent with the form of FZZ
one-point function\cite{FZZ} in the boundary Liouville Field Theory.
\end{abstract}

\vskip 2.25cm

\setcounter{footnote}{0}
\def\thefootnote{\fnsymbol{footnote}}
%

\resection{Introduction}
\label{intr}
Since the middle of 80's, there exist two independent approaches
to 2D Quantum Gravity: 
the continuous one called Liouville Gravity (LG) 
\cite{Polyakov,KPZ,David,DK} 
and the discrete one using the techniques of  Matrix Models (MM) \cite{BK,DS,GM,D,K,S,DFK}.
One may refer to more references in reviews in \cite{GM93,DFGZ}. 

Comparison of some quantities 
calculated in a number of particular models 
using both approaches 
confirms the expectation about 
their equivalence\cite {DFK,GL,BAlZ}.
The most easily checked quantities in MM are the $n$-point ``correlation numbers" 
\be
C_{k_1, \cdots, k_n} = \vev{O_{k_1} \cdots O_{k_n}} 
\ee
where $O_k$ is the integrated form of the local density (2-form operator) 
$\CO_k(X)$
\be
O_k = \int_\CM~\CO_k(X)
\ee
over the manifold $\CM$ which accommodates both ``matter" 
and the metric degrees of freedom localized at $X \in \CM$.
 
It is convenient to introduce a generating function of the correlation
numbers in Liouville gravity

\be
Z_{LG}(\{ \lambda_k\}) = \vev{e^{\sum_k \lambda_k O_k}}
\ee
which can be viewed as the partition function of the original 
theory perturbed by adding the fields $\CO_k(X)$ to the action density 
with the coupling constant $\lambda_k$. A similar partition function,
which depends on the parameters $t_k$, can be introduced in Matrix models.

However, a naive identification of observables and their correlation
numbers obtained from two approaches does not agree in general. 
The reason for this difficulty 
lies in the ambiguity of the choice 
of contact terms \cite{MSS,BZ}.
This ambiguity is equivalent to the freedom 
in the identification 
of the coupling parameters  $\lambda_k $ in  LG 
with the ones $t_k$ in MM 
of the same gravitational dimensions.
 
It was conjectured in \cite{MSS} 
that there exists a special choice of the contact terms 
in MM or, equivalently, 
the special transformation $t_k = t_k (\{\lambda_j\})$,
which ensures the coincidence of the partition functions in 
the Minimal Liouville gravity MLG(2,2p+1)
and in the $p$-critical one-matrix model OMM(p) 
for the random surfaces with the sphere topology. 
This  relation between the parameters  $t_k$ and $\lambda_k$  
was obtained in \cite{MSS} up to the  linear terms.
Its explicit form to all orders  was conjectured  in \cite{BZ} 
and checked up to the 4-th order \cite{BZ,GT}. 

The existence of a similar transformation 
for the general $(p_1,p_2)$ case remains an open problem,
and so much as the problem of the calculation 
of the correlation numbers and their comparison
on the surface with non-spherical topology. 

In this note, we consider the non-spherical topology,
namely, the fluctuating disk 
and propose the explicit expression for 
the generating function of the $n$-point bulk correlation numbers
on disk.
In our calculation, 
we use the  string equation \cite{BZ} in the conformal
frame.

Particularly, we compute the partition function 
and the one-point function
for the one-matrix model in this topology,
which were obtained earlier in \cite {MSS}
by means of a different way. 
We find our results consistent with the form of
the Fateev-Zamolodchikov-Zamolodchikov  
one-point function \cite{FZZ} 
in the boundary Liouville Field Theory.

In section \ref {OMM} 
we give a short summary of the known results on the OMM(p).
In section \ref{rmm}, the partition  function 
on the random disk with a fixed length of the boundary
and one-point function of the bulk operator in the OMM(p) 
are calculated. 
In section \ref{LG}, 
the corresponding answers are compared with the Liouville gravity 
results. 
Section \ref {Concl} is the conclusion and discussion. 

\resection{One-Matrix Model}
\label{OMM}
It was shown in \cite{BK,DS,GM} 
that the scaling partition function
of the OMM(p) is expressed in terms of the susceptibility
$u_*(t_k)$ which corresponds to the suitably chosen root  
of the string equation $P( u_*)=0$.
Here, $P(u)$ is the $p+1$-degree  polynomial of $u$ 
\be
\label{string-kdv}
 P(u,\mu, \{t_k\} ) 
 = u^{p+1} -\mu u^{p-1} + \sum_{k=1}^{p-1} t_k~u^{p-k-1}
\ee
where $\mu$ is the cosmological constant, 
and the parameters $t_k$ describe the relevant
deviations from the $p$-critical point. 

The singular (or universal) part of the partition function 
is expressed\cite{BZ} as 
\be
Z(\mu, \{t_k\} ) 
= {1\over2} \int_0^{u_*} du ~P^2(u,\mu, \{t_k\}) \,.
\ee
It was conjectured and partly checked in \cite{MSS,BZ} 
that there exists an analytic
transformation 
$t_k = t_k (\mu, \{\lambda_j\})$ 
compatible with the scale properties so that
the correlation numbers, defined by expanding 
the partition function into series 
of the new variables $\lambda_k $,
satisfy the same restrictions 
to be fulfilled in the minimal Liouville gravity. 
After this transformation
the polynomial $ P(u,t_k)$, as a function of the new
parameters $\lambda_k$, takes the form 
(up to  the  factor ${(p+1)!}/(2p-1)!! $) 
\be
\CP(u, \mu, \{\lambda_k\} ) 
\equiv P(u, \mu, \{t_k(\lambda)\} )
= u_0^{p+1}\CQ(u/u_0, \{\lambda_k \} )
\ee
where  
\bea
\CQ(x, \{\lambda_k \} ) 
&=& \sum\limits_{n=0}^{\infty}\sum\limits_{k_{1}...k_{n}=1}^{p-1}
 \frac{\lambda_{k_1}\lambda_{k_2}\cdots \lambda_{k_n} }{n!} 
  (\frac{d}{dx})^{n-1}L_{p-\sum k_{i}-n}(x),
\\ 
(\frac{d}{dx})^{-1} L_p (x) &\equiv& 
\int_0^x L_(y) dy = \frac{L_{p+1}(x) - L_{p+1}(x)}{2p+1} \,.
\eea
$u_*(\lambda)$ is the suitably chosen root of $\CP(u,\lambda_k)$
and reduces to $u_0$ if $\lambda_k=0$ as given in Eq.~(3.26) of \cite{BZ}
\be
 u_0= \sqrt{\frac{2 (2p-1)}{ p(p-1)}\mu}\,. 
\ee

The partition function in the new variables is rewritten as 
\be
\CZ(\mu, \{\lambda_k\} ) 
= {1\over2} \int_0^{u_*}  \CQ^2(u/u_0,\lambda_k) du 
\ee
and is considered as the generating function 
of the correlation numbers
in the Liouville frame.

It is important to remember that these new correlation numbers
are equal to the coefficients of the expansion of the partition function
$\CZ(\mu, \{\lambda_k\})$ around the point $\lambda_k=0$ which
does not coincide with the point $ t_k=0 $.
The last one was  used to define the correlation numbers in the 
KdV frame.
We will refer to the new choice of the point of the expansion
as the Liouville (or the conformal) background
following \cite{MSS}.

\resection{ Bulk one-point function on disk }
\label{rmm}

Now we consider the fluctuating disk with the boundary length $L$.
It was shown in \cite{GM,Banks} that the partition function of the random disk with a finite boundary is given by 
\bea
\label{kdv-pf}
Z_{B}(\mu, L, t_1,\cdots, t_{p-1} ) 
=\frac{1}{\sqrt{L}} 
\int_{t_{p-1}}^\infty dy~ e^{-L\,u(y)} 
\eea
where  $u(y)$ stands for the solution of the string equation 
in the KdV frame 
\bea
\label{streq-kdv}
 P(u, \mu, t_1, \cdots, t_{p-2}, y) = u^{p+1} -\mu u^{p-1} 
 + \sum_{k=1}^{p-2} t_k~u^{p-k-1}+ y = 0\,.
\eea

The disk partition function $Z_{B}(\mu, L, {t_k=0})$
in the KdV frame was computed by Moore, Seiberg, and Staudacher.
The answer is given in Eqs.~(4.4) and (4.5) of \cite{MSS}.

However, to compare the results of MM 
with the ones of MLG on disk we need to use the Liouville frame.
Therefore, we will start from the formula
\bea
 \CZ_{B}(\mu, L,  \{\lambda_k \} ) 
 =\frac1{\sqrt{L}} 
\int_{\lambda_{p-1}}^\infty dy ~ e^{-L\,  u(y)} 
\eea 
where  $u(y)$ stands for the solution of the string equation 
in the  Liouville frame 
\be
\CP(u,\mu,\lambda_1,\cdots,\lambda_{p-2},y)=0\,.
\ee
We can compute the partition function and the $n$-point functions
on disk using the string equation after changing the integration 
variable $y$ to $u$. For this we use the relation
\be
\frac{dy}{du}=-\frac{d\CP(u,\mu, \{\lambda_k\})}{du}
\ee
and obtain 
\bea
\CZ_B(\mu, L, \{\lambda_k\}) 
&=& -\frac { u_0^{p+1}}{\sqrt{L}} 
\int_{x_*}^\infty dx ~ \frac{dQ(x,\{\lambda_k\})}{dx} e^{-L u_0 x} 
\nn\\
&=& u_0^{p+2} {\sqrt{L}} 
\int_{x_*}^\infty dx ~ Q(x,\{\lambda_k\}) e^{-L u_0 x} 
\label{macroloop2}
\eea 
where the integration by parts is performed and $x_*$ is the corresponding root 
of the renormalized string equation $ Q(x_*,\{\lambda_k\})=0$.
The formula (\ref{macroloop2}) is the expression for the generating function
of the correlation numbers on the disk in the Liouvile frame.
The disk partition function is obtained when $\lambda_k=0$
\bea
\CZ_B(\mu, L,\{ \lambda_k\}=0)
 =\frac {u_0^{p+1}} {\sqrt L}  \, 
 \int_1^\infty  dx  L_p(x)  e^{- L u_0  x} \,.
 \label{z} 
\eea
The bulk one-point function is given as 
\bea 
\vev{O_k}_{L}
&=& \frac{\partial}{\partial\lambda_k} \CZ_B(\mu, L, \lambda_1,\cdots,\lambda_{p-1})\Big|_{\{\lambda_i\}=0} 
\nn\\ 
&=&  u_0^{p-k}{\sqrt{L}} 
\int_{x_*}^\infty dx ~ L_{p-1-k}(x) e^{-L u_0 x} 
\label{op}
\eea

One can evaluate the expressions 
using Eqs.~(\ref{z}) and (\ref{op}), 
and the relation \cite{BZ} 
between the Legendre polynomials and the Macdonald function
of a half-integer order
\be
\int_{1}^\infty dx ~ L_n(x) e^{-p x} 
=\sqrt{ \frac{2}{ \pi p}}  K_{n+1/2} (p) \,.
\ee
The explicit results are  
\bea
\CZ_B(\mu, L, \lambda_k=0) 
 &=& 
 \sqrt{\frac 2  \pi }\frac {u_0^{p+1/2}}L  K_{p+1/2}(u_0 L) 
\\
\vev{O_k}_{L} 
&=& 
\sqrt{ \frac2 \pi}   u_0^{p-k-1/2} K_{p-k-1/2} (u_0 L)\,.
\label{OL}
\eea
These relations coincide with formulas (4.19) and (4.24) in \cite{MSS}
where they were obtained in a different way.

\resection{Comparison with Boundary Liouville field theory}
\label{LG} 

Noting that the Liouville gravity partition function 
with the fixed boundary length $\ell$, 
$Z_{BLG} (\mu,\ell, \{\lambda_k\})$ 
is related to the one 
with the fixed boundary cosmological constant $\mu_B$,
$\CZ(\mu,\mu_B, \{\lambda_i\})$ 
by the inverse Laplace transform 
\be
Z_{BLG}(\mu, \ell, \{\lambda_k\}) 
= \ell \int_{\uparrow} 
\frac{d\mu_B}{2\pi i} ~ e^{\mu_B \ell}
~Z_{BLG}(\mu,\mu_B, \{\lambda_k\}) \,,
\ee
where the contour $\uparrow$ goes along the imaginary 
axis to the right from all the singularities 
of the integrand. 

One can express the one-point correlation number 
with fixed $\ell$ 
in terms of the one with fixed $\mu_B$ as 
\be
\vev{O_k}_{\ell}= 
\ell \int_{\uparrow} 
\frac{d\mu_B}{2\pi i} ~ e^{\mu_B \ell} ~
\vev{O_k}_{\mu_B}
\ee
where 
\be
\vev{O_k}_{\mu_B} = 
\frac{\partial}{\partial \lambda_k} 
Z_{BLG}(\mu,\mu_B, \{\lambda_i\})\Big|_{\{\lambda_i\}=0}\,.
\ee

The bulk one-point function in the Boundary Liouville field theory
is obtained in \cite{FZZ} 
\be
U(\alpha|\mu_B) = 
\frac2b (\pi \mu \gamma(b^2))^{(Q- 2\alpha)/(2b)}
\Gamma(2 b \alpha-b^2) ~\Gamma(\frac{2\alpha}b -\frac1{b^2})
~\cosh((2 \alpha -Q) \pi s)  
\ee
with 
\be
\cosh^2(\pi bs)= \frac{\mu_B^2}\mu ~\sin (\pi b^2 ) \,.
\ee
Using the Laplace transform, one gets
\bea
U(\alpha|\mu_B) 
&=& \int_0^\infty \frac{d\ell}\ell 
e^{-\mu_B \ell}~ W_\alpha(\ell) 
\\
W_\alpha(\ell)&=& 
\frac2b (\pi \mu \gamma(b^2))^{(Q-2\alpha)/(2b)} 
\frac{ \Gamma(2 b \alpha-b^2)}{\Gamma(1+ \frac1{b^2}-\frac{2\alpha}b)}
K_{(Q-2\alpha)/b} (\kappa \ell) 
\label{one-point-fixed-ell}
\eea
with 
\be
\kappa^2 =\frac  \mu{\sin (\pi b^2)}\,.
\ee
Noting that 
$1/{b^2} = p+ 1/2 $ 
and putting $\alpha_k  = (k+2) b/2\,$, ~one has 
\be
\frac{Q-2\alpha_k}b = p-k -\half 
\ee
which demonstrates the consistency 
between the matrix model and the Liouville gravity on disk since $W_{\alpha_k}(\ell)$ has the same dependence on  $\mu$
and $\ell$  as the one in $\vev{O_k}_L$ in Eq.~(\ref{OL})
up to the renormalization of $L$. 
\resection{Conclusion and Discussion}
\label{Concl}

The result (\ref{macroloop2}) 
gives an expression for the generating function 
of the correlation numbers of 
the  $p$-critical one-matrix model on disk in the Liouvile frame.  
To confirm this proposal, we explicitly 
compare the one-point correlation numbers  obtained from (\ref{macroloop2}) 
against the FZZ one-point function 
Eq.~(\ref{one-point-fixed-ell})
in the Boundary Liouville 
field theory and find the proposal consistent. 

One may further evaluate the bulk $n$-point functions
using (\ref{macroloop2}).  
However, one cannot check their correctness at this
moment, since the corresponding answers
for MLG are not known yet.   

Another interesting problem in both MM and MLG 
will be the computation of bulk-boundary correlation numbers 
and boundary correlation numbers on disk \cite{Hoso}.

\section{Aknowledgements} 
Authors are indebted to V. Belavin, G. Ishiki, Y. Ishimoto, M. Lashkevich, and H. Shin for useful discussions and comments. 
A.B. is grateful to the Center for Quantum Spacetime
of Sogang University where important part of this work 
was performed.
The research of A.B. was held within the bounds
of Federal Program "Scientific and Scientific-Pedagogical personnel
of innovational Russia", RFBR initiative interdisciplinary
project 09-02-12446-ofi-m and RFBR-CNRS project PICS-09-02-91064. 
C.R. was partially supported 
by National Research Foundation of Korea (NRF)
funded by the Korea government(MEST)
with grant number 2005-0049409 and R11-2005-021.

\vskip 1cm



\begin{thebibliography}{99}
\raggedright
\parskip 1pt
%
%
\bibitem{Polyakov}
A.\ Polyakov, 
``Quantum geometry of bosinic strings'', 
Phys.\ Lett.\ {\bf B103} (1981) 207.
%
\bibitem{KPZ}
V.\ Knizhnik,A.\ Polyakov and A.\ Zamolodchikov,
``Fractal Structure of 2D quantum gravity'',
Mod.\ Phys.\ Lett.\  {\bf A3} (1988) 819. 
%
\bibitem{David}
F.\ David,
``Conformal Field Theories Coupled to 2-D Gravity in the Conformal Gauge'', 
 Mod.\ Phys.\ Lett.\  {\bf A3} (1988) 1651.
%
\bibitem{DK}
J.\ Distler and H.\ Kawai,   
``Conformal Field Theory and 2D Quantum Gravity Or Who's Afraid of Joseph Liouville?'', 
Nucl.\ Phys.\ {\bf B321} (1989) 509.
%
\bibitem{BK}
E.\ Brezin and V.\ Kazakov,
``Exactly Solvable Field Theories Of Closed Strings'',
Phys.\ Lett.\ {\bf B236} (1990) 144. 
%
\bibitem{DS}
M.\ Douglas and S.\ Shenker,
``Strings in Less Than One-Dimension'', 
Nucl.\ Phys.\ {\bf B335} (1990) 635.
%
\bibitem{GM}
D.\ Gross and A.\ Migdal,
``Nonperturbative Two-Dimensional Quantum Gravity'',
 Phys.\ Rev.\ Lett.\ {\bf 64} (1990) 127;
 ``A Nonperturbative Treatment Of Two-Dimensional Quantum Gravity'',
Nucl.~Phys.~{\bf B340} (1990) 333. 
%
\bibitem{D}
M.\ Douglas,
``Strings In Less Than One-Dimension And The Generalized K-D-V Hierarchies'', 
Phys.~Lett.~{\bf B238} (1990) 176. 
%
\bibitem{K}
V.\ Kazakov,
``The Appearance of Matter Fields from Quantum Fluctuations of 2D Gravity'',
Mod.\ Phys.\ Lett {\bf A4} (1989) 2125.
%
\bibitem{S}
M.\ Staudacher
``The Yang-Lee singularity on a dynamical planar random surface'',
 Nucl.~Phys.~{\bf B336} (1990) 349. 
%
\bibitem{DFK}
P.\ Di Francesco and D.\ Kutasov,
``World sheet and space time physics in two dimensional 
(super) string theory'',
Nucl.~Phys.~{\bf B375} (1992) 119.
%
\bibitem{GM93}
P.\ Ginsparg and G.\ Moore,
``Lectures on 2-D gravity and 2-D string theory (TASI 1992)'',
{\tt  arXiv:hep-th/9304011}
%
\bibitem{DFGZ}
P.\ Di Francesco,P.\ Gisparg and J.\ Zinn-Justin,
``2-D gravity and random matrices'',
 Phys.~Rep.~{\bf 254} (1995) 1.  
%
\bibitem{GL}
M.\ Goulian and M.\ Li,
``Correlation functions in Liouville theory'',
Phys.~Rev.~Lett.~{\bf 66} (1991) 2051.
%
\bibitem{BAlZ}
A.\ Belavin and Al.\ Zamolodchikov,
``Moduli integrals,ground ring and four-point function 
in minimal Liouville gravity'',
Theor.~Math.~Phys.~{\bf  147} (2006) 729; 
{\tt arXiv:hep-th/0510214}
 %
 \bibitem{MSS}
 G.~Moore, N.~Seiberg, M.~Staudacher, 
``From loop to states in 2D quantum gravity'',
Nucl.~Phys.~{\bf B362} (1991) 665.
%
\bibitem{BZ}
A.A.\ Belavin and A.\ Zamolodchikov, 
``On correlation numbers in 2D minimal gravity and matrix models'',
Jour.~Phys.~{\bf A42} (2009) 304004; 
{\tt arXiv: 0811.0450[hep-th]}  .
%
\bibitem{GT}
G.\ Tarnopolsky,
``Five-point Correlation Numbers in One-Matrix Model'',
{\tt arXiv: 0912.4971[hep-th]}
%
\bibitem{Banks}
T.\ Banks, M.\ Douglas, N.\ Seiberg and S.\ Shenker,
``Microscopic and macroscopic loops in non-perturbative 
two dimensional gravity'',
Nucl.~Phys.~{\bf B238} (1990) 279. 
 %
\bibitem{FZZ}
V.\ Fateev, A.\ Zamolodchikov and  Al.\ Zamolodchikov, 
``Boundary Liouville field theory. 1. Boundary state and boundary two point function'', {\tt hep-th/0001012}. 

\bibitem{Hoso}
K. Hosomichi, ``Minimal Open Strings", 
JHEP {\bf 0806}:029 (2008); 
 J.-E. Bourgine and K. Hosomichi, 
``Boundary operators in the O(n) and RSOS matrix models",  
JHEP {\bf 0901}:009 (2009);
Jean-Emile Bourgine, Kazuo Hosomichi andIvan Kostov,
``Boundary transitions of the O(n) model on a dynamical lattice", 
{\tt arXiv: 0910.1581[hep-th]} 
 
\end{thebibliography}
\end{document}